%% file: main.tex
\documentclass[letterpaper, 10 pt, conference]{cssconf}
\IEEEoverridecommandlockouts 
\usepackage{url} %
\usepackage{amsmath} %
\usepackage{amsfonts}%
\usepackage{bm}
\usepackage{algorithm}
\usepackage{adjustbox} %
\usepackage{tikz}
\usetikzlibrary{arrows.meta}
\usepackage{multirow}
\usepackage{booktabs} 
\usepackage{csvsimple} 

\newtheorem{theorem}{Theorem}
\newtheorem{rem}{Remark}

\title{Inverse Learning in $2\times2$ Games: From Synthetic Interactions to Traffic Simulation}

\author{Daniela Aguirre Salazar, Firas Moatemri and Tatiana Tatarenko
\thanks{The authors are with the Department of Control Theory and Intelligent Systems, TU Darmstadt, Germany. E-mails: {\tt daniela.aguirre@tu-darmstadt.de},
{\tt firas.moatemri@stud.tu-darmstadt.de},
{\tt tatiana.tatarenko@tu-darmstadt.de}.}}

\begin{document}
\maketitle
\thispagestyle{empty}
\pagestyle{empty}

\begin{abstract}
Understanding how agents coordinate or compete from limited behavioral data is central to modeling strategic interactions in traffic, robotics, and other multi-agent systems. In this work, we investigate the following complementary formulations of inverse game-theoretic learning: (i) a \emph{Closed-form Correlated Equilibrium Maximum-Likelihood} estimator (CE-ML) specialized for $2\times2$ games; and (ii) a \emph{Logit Best Response Maximum-Likelihood} estimator (LBR-ML) that captures long-run adaptation dynamics via stochastic response processes. Together, these approaches span the spectrum between static equilibrium consistency and dynamic behavioral realism. We evaluate them on synthetic “chicken-dare” games and traffic-interaction scenarios simulated in SUMO, comparing parameter recovery and distributional fit. Results reveal clear trade-offs between interpretability, computational tractability, and behavioral expressiveness across models.
\end{abstract}

\section{Introduction}

Game theory provides a principled foundation for modeling strategic interactions among multiple decision-making agents. Traditional approaches assume that each player's utility function is known and aim to predict equilibrium behaviors accordingly. In contrast, the \emph{inverse} perspective seeks to infer these latent utilities from observed behavior, explaining why agents act as they do rather than merely describing the resulting equilibria. This \emph{inverse game-theoretic learning} paradigm is especially relevant in domains such as traffic, robotics, and human-machine interaction, where preferences and coordination mechanisms are implicit and cannot be directly measured~\cite{Chandra2025MAIRLcrowdsIROS, KuleshovSchrijvers2015InverseGameTheory, pmlr-v97-yu19e}.

A central challenge in inverse learning lies in selecting the behavioral or equilibrium concept that best captures observed coordination. The classical \emph{Nash Equilibrium} (NE) assumes fully independent strategy choices and perfect rationality, but it is both computationally intractable in general games, PPAD-complete~\cite{daskalakis2009complexity}, and often inconsistent with empirical data. In many multi-agent settings, including driving, agents exhibit partial coordination or correlated behavior that deviates from strict Nash predictions~\cite{camerer2003behavioral,nyarko2002belief}.

To address these limitations, the \emph{Correlated Equilibrium} (CE)~\cite{aumann1974subjectivity} offers both tractable computation~\cite{PapadimitriouRoughgarden2008} and richer behavioral realism. CE relaxes the independence assumption of Nash Equilibrium by allowing agents’ actions to be conditionally correlated through shared signals or learned expectations~\cite{DuffyLaiLim2017CoordinationViaCorrelation}. Moreover, regret-based adaptive dynamics are known to converge to the CE set~\cite{hart2000simple}, providing a natural bridge between equilibrium theory and repeated-play learning, which is particularly relevant in domains like traffic, where drivers continually adjust to each other’s behavior.
Building on this foundation, recent work has explored \emph{inverse} methods for estimating latent utilities consistent with observed correlated play. Among these, the \emph{Inverse Correlated Equilibrium (ICE)} framework~\cite{bestick2013inverse} provides a principled baseline: it infers utilities that make the observed empirical distribution an approximate correlated equilibrium. However, the ICE-based utility learning in~\cite{bestick2013inverse} sidesteps principled statistical estimation of the latent distributions. The proposed linear programming approach merely enforces CE constraints in expectation under the empirical distribution rather than modeling and fitting the correlated equilibrium distribution itself.

Another solution concept that aims to capture more realistic strategic behavior is the \emph{Quantal Response Equilibrium (QRE)}~\cite{MCKELVEY19956}, which models bounded rationality by allowing players to make probabilistic rather than perfectly rational choices.
Inverse QRE frameworks have been used to analyze conflict interactions in traffic domains, typically represented as $2\times 2$ normal-form games~\cite{ArbisDixit2019LaneChangingConflicts, ZhangFricker2021ConflictRisksPMI}.
However, these works do not provide a theoretical justification for the specific form of the joint action distribution implied by QRE.
Moreover, computing a QRE is as hard as computing a Nash equilibrium~\cite{MCKELVEY19956} and is therefore PPAD-complete, making inverse QRE learning generally intractable.

To address these limitations, we introduce two novel inverse learning formulations that broaden both the scope and interpretability of equilibrium-based inference in~$2~\times~2$ normal-form games, a structure commonly encountered in practical domains such as traffic interactions, security scenarios, and human-robot decision-making~\cite{ArbisDixit2019LaneChangingConflicts, CostaGomesCrawfordBroseta2001, RazinFeigh+2021+481+502}.
The first, the \emph{Closed-Form Correlated Equilibrium Maximum-Likelihood} (CE-ML) estimator, exploits the analytical geometry of the correlated equilibrium polytope in 
$2\times2$ games~\cite{CalvoArmengol2003_CE2x2} to enable direct estimation of the equilibrium distribution and efficient parameter fitting.
The second, the \emph{Logit Best Response Maximum-Likelihood} (LBR-ML) framework, departs from static equilibrium reasoning and models the long-run distribution of play as the stationary state of simultaneous logit best response dynamics~\cite{Blume1993}.
Together, these formulations span a spectrum of modeling choices, from expectation-based equilibrium consistency (ICE) to analytically tractable estimation (CE-ML) and dynamically grounded behavioral realism with bounded rationality (LBR-ML).

We empirically evaluate these three approaches in two representative domains. First, in synthetic “chicken-dare” games with known ground-truth utilities~\cite{bestick2013inverse}, we measure parameter recovery accuracy and equilibrium consistency. Second, in a SUMO-based traffic simulation~\cite{SUMO2018}, we examine whether the inferred utilities reproduce realistic yielding and proceeding patterns in uncontrolled intersections. This evaluation demonstrates both methodological differences and practical implications for modeling strategic behavior in interactive systems.


\section{Problem Formulation and Equilibrium Concepts}
\label{sec:equilibrium-concepts}
\subsection{Inverse Game-Theoretic Learning}
We consider a finite normal-form game $\Gamma_N^w = \Gamma\{N, \{A_i\}, \{u^{w_i}_i\}\}$ with players from the set $[N]=\{1,\dots,N\}$.  
Each player $i\in [N]$ has an action set $A_i$, and the joint action set is  
$A=\prod_{i\in [N]}A_i$. We denote by $n$ the cardinality of the set $A$. For a joint action $a=(a_i,a_{-i})\in A$, $a_i$ denotes the action of the player $i$, while 
$a_{-i}$ denotes the actions of all players except for $i$.  

Each player receives a utility $u^{w_i}_i(a)\in\mathbb{R}$, assumed to be linear in a known feature representation $\phi_i(a)\in\mathbb{R}^{d_i}$:
\begin{equation}
\label{eq:utility-linear}
u^{w_i}_i(a)=\phi_i(a)^\top w_i,
\end{equation}
where $w_i\in\mathbb{R}^{d_i}$ are player-specific weights to be estimated.  
The utility expression~\eqref{eq:utility-linear} captures interpretable components of decision-making, for example, risk, comfort-related cost, or social preference, through features $\phi_i(a)$ and their relative importance encoded by $w_i$. 
Classical analysis assumes known utilities $\{u^{w_i}_i\}$ and intends to predict a stable state (so called equilibrium) in $\Gamma_N^w$.  
\textbf{Inverse game-theoretic learning} reverses this logic: \emph{given observed joint actions
$\mathcal{D}=\{a^{(t)}\}_{t=1}^T$, we seek utility parameters $w = (w_1,\ldots, w_N)\in\mathbb R^{Nd}$ that make the observations consistent with a definite equilibrium strategy in the game}.

The next section introduces several equilibrium concepts relevant to finite normal-form games.

\subsection{Equilibrium Concepts}
A mixed strategy for player $i$ in the game $\Gamma_N^w$ is a probability distribution  
$\sigma_i\in\Delta(A_i)$, where $\Delta(A_i)$ is the set of all possible probability distributions over  $A_i$, and a joint mixed strategy is  
$\sigma=(\sigma_i)_{i\in [N]}\in\Delta(A) = \prod_{i\in [N]}\Delta(A_i)$. Unless stated otherwise, $\sigma$ is assumed to be a \emph{row vector} of dimension equal to the cardinality of $A$. 
The expected utility of player $i$ under $\sigma$ is, thus, defined by
\begin{equation}
\tilde u^{w_i}_i(\sigma) 
=\sum_{a\in A}\sigma[a]\,u^{w_i}_i(a),
\end{equation}
where $\sigma[a]=\prod_{j\in [N]}\sigma_j[a_j]$ denotes the probability assigned by the joint mixed strategy to the pure action profile $a=(a_1,\dots,a_N)$.
Throughout the paper, we use the notation $\sigma[\cdot]$ to refer to probabilities of specific joint actions,
whereas $\sigma(\cdot)$ denotes the parametric dependence of the strategy profile on model/games parameters such as, for example, the vector $w$ to be estimated. When both aspects appear simultaneously, we write $\sigma(\cdot)[\cdot]$.\footnote{For example, 
$\sigma(w)[a]$ denotes the probability of action $a$ under the mixed strategy $\sigma$ given the utility function parameterization $w$.}

The classical \emph{mixed strategy Nash Equilibrium} (NE) requires that each player's mixed strategy is a best response to the others'.  
Formally, $\sigma_\text{NE}^\star(w)$ is an NE if, for all $i$ and all $\sigma_i\in\Delta(A_i)$,
\[
\tilde u^{w_i}_i\!\big(\sigma_{\text{NE},i}^{\star}(w),\;
\sigma_{\text{NE},-i}^{\star}(w)\big)
\;\ge\;
\tilde u^{w_i}_i\!\big(\sigma_i,\;
\sigma_{\text{NE},-i}^{\star}(w)\big).\footnote{We use $\sigma_{-i}$ to denote the joint mixed strategy of all players except for the player $i$}
\]
Computing NE is PPAD-complete in general~\cite{daskalakis2009complexity}, and empirical studies show that human play often violates its independence assumption~\cite{camerer2003behavioral,nyarko2002belief}, motivating equilibrium notions that allow correlated or adaptive coordination. In this paper, we focus on two examples of such notions: \emph{Correlated Equilibrium} and \emph{stationary state of logit best response dynamics}.

\paragraph{Notion I: Correlated Equilibrium} The \emph{Correlated Equilibrium} (CE)~\cite{aumann1974subjectivity} generalizes the Nash Equilibrium by allowing correlation among players’ actions through a joint distribution, typically induced by an external correlation device introduced to the system.
A distribution $\hat{\sigma}^\star(w)$ is a CE in $\Gamma_N^w$ if, for all players $i$ and all deviations $a_i'\in A_i$,
\begin{align}\label{eq:CElin}
    \sum_{a\in A}\hat{\sigma}^\star(w)[a_i,a_{-i}]\,\big[u^{w_i}_i(a_i,a_{-i})-u^{w_i}_i(a_i',a_{-i})\big] \ge 0.
\end{align}
CE retains convexity and can be computed efficiently as a linear program, while the correlation device provides the probabilistic mechanism through which equilibrium joint actions can be estimated from data.

\paragraph{Notion II: Stationary State of Logit Best Response Dynamics} Alternatively, behavioral models based on the \emph{Logit Best Response} (LBR)~\cite{Blume1993} relax the assumptions of perfect rationality and centralized correlation by allowing stochastic choice governed by player-specific rationality parameters $\lambda_i \ge 0$.
Let $\lambda=(\lambda_1,\ldots,\lambda_N)$ denote the vector of rationality levels.  
Enumerate the joint action set as $A=\{a(1),\ldots,a(n)\}$, and let $a_i(l)$ denote the $i$-th player's component of $a(l)$.  
Given the current joint action $a(k)\in A$, the LBR induces the mixed strategy
\begin{align}
\label{eq:lbr2}
\sigma(\lambda,w)[a(l)|a(k)]
  &= \prod_{i=1}^N \sigma_i(\lambda_i, w_i)[a_i(l)|a_{-i}(k)], \cr &\qquad\qquad\qquad  a(l)\in A,\cr
\sigma_i(\lambda_i, w_i)[a_i(l)&|a_{-i}(k)]\cr
  = &\frac{\exp\!\big(\lambda_i\,u^{w_i}_i(a_i(l),a_{-i}(k))\big)}
          {\sum_{a_i'\in A_i}\exp\!\big(\lambda_i\,u^{w_i}_i(a_i',a_{-i}(k))\big)}.
\end{align}

Higher values of $\lambda_i$ indicate that player~$i$ behaves more deterministically (i.e., more ``rationally'').  
At $\lambda_i=0$, choices are uniform over $A_i$, whereas as $\lambda_i\to\infty$, $\sigma_i(\lambda_i, w_i)[\cdot|a_{-i}(k)]$ concentrates on best replies to $a_{-i}(k)$ (assuming a unique best reply or a fixed tie-breaking rule).

It is worth noting that to follow LBR, players do not need any coordinator or correlating device, as it is the case for the CE notion. Each player needs to observe the last realized joint action $a(k)$ and to know the corresponding payoffs $u^{w_i}_i(\cdot,a_{-i}(k))$.  
Hence, decisions depend only on the most recent joint action, which makes the process \emph{myopic} and \emph{memoryless}~\cite{Blume1993, Tatarenko2017GameTheoreticLearning}.  
This distinguishes the present LBR dynamics from the \emph{quantal response equilibrium} (QRE), where logit responses are taken with respect to expected utilities under mixed strategies and, thus, require the players to exchange their mixed strategies.

If players follow the LBR rule simultaneously at every stage, the induced process on the finite set $A$ is a time-homogeneous Markov chain with transition matrix
\begin{align}
\label{eq:transition}
P(\lambda, w) = (p_{kl}(\lambda,w))_{k,l=1}^n,
\end{align}
where $p_{kl}(\lambda,w) = \sigma(\lambda,w)[a(k)|a(l)]$ in defined by~\eqref{eq:lbr2}.
That is, $p_{kl}(\lambda,w)$ is the probability of transitioning from joint action $a(k)$ to $a(l)$.  
Let $\sigma^t\in\Delta(A)$ denote the distribution over the joint actions at time~$t$. Then the evolution of the joint mixed strategy is
\begin{align}\label{eq:LD}
\sigma^{t+1} = \sigma^t\, P(\lambda, w),
\end{align}
given that the players start from some initial $\sigma^0\in\Delta(A)$.

Denote by $\sigma^\star(\lambda,w)$ the stationary distribution of $P(\lambda, w)$, that is,
\begin{align}\label{eq:stationary}
\sigma^\star(\lambda,w)\, P(\lambda, w) = \sigma^\star(\lambda,w).
\end{align}
The stationary distribution exists due to the row stochasticity of $P(\lambda, w)$. 

If $0\le\lambda_i<\infty$ for all $i$ and utilities are finite, then each term in~\eqref{eq:lbr2} is strictly positive, so $P(\lambda, w)$ is a \emph{primitive} stochastic matrix.  
By the Perron-Frobenius theorem for finite Markov chains, $P(\lambda, w)$ admits a \emph{unique} stationary distribution and the chain \emph{converges} to it exponentially fast~\cite{Seneta2006}. Thus, we obtain the following result.
\begin{theorem}\label{th:st}
Let the players in $\Gamma_N^w$ choose LBR strategies simultaneously over time according to~\eqref{eq:LD} with finite rationality levels $\lambda_i<\infty$ and starting by any mixed strategy $\sigma^0\in\Delta(A)$.  
Then there exists a unique stationary distribution $\sigma^\star(\lambda,w)$  such that
$
\lim_{t\to\infty} \sigma^t = \sigma^\star(\lambda,w).
$
\end{theorem}

Theorem above implies that the dynamics stabilizes around the unique stationary distribution $\sigma^\star(\lambda,w)$ of $P(\lambda,w)$.  
This $\sigma^\star(\lambda,w)$ is referred to as the \emph{stationary state of logit best response dynamics} in $\Gamma_N^w$, and it serves as the reference equilibrium distribution in our second inverse game-theoretic learning framework.

\section{Methodology for Inverse Game-Theoretic Learning in 2$\times$ 2 Games}

In this section, we describe in details two frameworks for inverse game-theoretic learning:
(i) a \emph{Closed-form Correlated Equilibrium Estimator} for $2\times2$ games (CE-ML),
and (ii) a \emph{Logit Best Response Learning} (LBR-ML) approach based on the stationary distribution of logit response dynamics.
Each method uses the specific equilibrium solution notion from the previous section to connect the observed behavioral data with latent utilities parameterized as in~\eqref{eq:utility-linear}.


To be more specific, let us consider a game $\Gamma^w_2$, where \([2]=\{1,2\}\) is the set of players, Player~1 is the row player and Player~2 is the column player (see Table~\ref{tab:2x2-payoffs}).
For \(i\in[2]\), let \(A_i=\{a_i^1,a_i^2\}\) be the action set of Player \(i\).  
The joint action set is \(A = A_1 \times A_2\), which we enumerate as
$
A=\{\,a(1)=(a_1^1,a_2^1),\; a(2)=(a_1^1,a_2^2),\; a(3)=(a_1^2,a_2^1),\; a(4)=(a_1^2,a_2^2)\,\}.
$
For each player \(i\), the payoff at a joint action \(a(l)\) is denoted by \(u^{w_i}_i(a(l))\).  
As before, \(w_1\in\mathbb{R}^d\) and \(w_2\in\mathbb{R}^d\) are the utility parameter vectors to be estimated (see the linear utility specification in~\eqref{eq:utility-linear}).


\begin{table}[H]
  \centering
  \caption{Two-player \(2\times2\) game: (a) payoff matrix \((u^{w_i}_1,u^{w_i}_2)\).}
  \label{tab:2x2-payoffs}
  \renewcommand{\arraystretch}{1.2}
  \begin{tabular}{c|cc}
    \hline
    & \(a_2^1\) & \(a_2^2\) \\
    \hline
    \(a_1^1\)
      & \((u^{w_1}_1(a(1)),\,u^{w_2}_2(a(1)))\)
      & \((u^{w_1}_1(a(2)),\,u^{w_2}_2(a(2)))\) \\
    \(a_1^2\)
      & \((u^{w_1}_1(a(3)),\,u^{w_2}_2(a(3)))\)
      & \((u^{w_1}_1(a(4)),\,u^{w_2}_2(a(4)))\) \\
    \hline
  \end{tabular}
\end{table}

\subsection{Approach 1: Closed-form 2$\times$2 CE Maximum-Likelihood (CE-ML)}
It has been shown in \cite{CalvoArmengol2003_CE2x2} that for the class of \(2\times 2\) normal-form games, the correlated-equilibrium (CE) set admits a compact closed-form characterization defined by four linear inequalities~\eqref{eq:CElin} and the simplex constraints. Geometrically, this feasible region forms a convex polytope that can be expressed as the convex hull of a small number of extreme points corresponding to feasible pure or mixed joint-action profiles.

We follow the next assumption introduced in~\cite{CalvoArmengol2003_CE2x2}, which allows for the full characterization of the CE set in 2$\times$2 normal-form games. 

\noindent\textbf{Assumption (no dominated strategies).}
No action of either player is strictly dominated; in particular, 
$u^{w_1}_1(a(1)) \neq u^{w_1}_1(a(3))$, 
$u^{w_1}_1(a(4)) \neq u^{w_1}_1(a(2))$, 
$u^{w_2}_2(a(1)) \neq u^{w_2}_2(a(2))$, 
$u^{w_2}_2(a(4)) \neq u^{w_2}_2(a(3))$.

Based on the payoff matrix in Table~\ref{tab:2x2-payoffs} 
$(u^{w_1}_1(a(l)),\,u^{w_2}_2(a(l)))_{l\in\{1,2,3,4\}}$ 
and taking the assumption above into account, let us define
\begin{align*}
&\alpha^{w_1}=
\frac{\lvert u^{w_1}_1(a(1)) - u^{w_1}_1(a(3)) \rvert}
     {\lvert u^{w_1}_1(a(4)) - u^{w_1}_1(a(2)) \rvert},
\cr
&\beta^{w_2}=
\frac{\lvert u^{w_2}_2(a(1)) - u^{w_2}_2(a(2)) \rvert}
     {\lvert u^{w_2}_2(a(4)) - u^{w_2}_2(a(3)) \rvert}.
\end{align*}

\begin{theorem}\label{th:1}
Consider a (strict) coordination game \(\Gamma^w\) with
$u^{w_1}_1(a(1)) > u^{w_1}_1(a(3))$, 
$u^{w_1}_1(a(4)) > u^{w_1}_1(a(2))$,
$u^{w_2}_2(a(1)) > u^{w_2}_2(a(2))$, 
$u^{w_2}_2(a(4)) > u^{w_2}_2(a(3))$.
Then the set of correlated equilibria in \(\Gamma^w_2\) is a polytope with five vertices $\{\hat{\sigma}^\star_{(v)}(w)\}_{v=1}^5$ listed in Table~\ref{tab:ce-vertices} \footnote{For notation simplicity, in Table~\ref{tab:ce-vertices},  we omitted the dependence on $w_1$ and $w_2$ in $\alpha$ and $\beta$ respectively.}, where we denote the probability of the joint action $a(l)$, $l=1,2,3,4$, simply by
$\hat{\sigma}^\star(w)[a(l)]$.
\end{theorem}
See \cite{CalvoArmengol2003_CE2x2} for the proof.

\begin{table}[H]
  \centering
  \caption{Five vertices of the correlated-equilibrium polytope (strict coordination case), as characterized in~\cite{CalvoArmengol2003_CE2x2}.}
  \label{tab:ce-vertices}
  \begingroup
  \setlength{\tabcolsep}{3pt}
  \renewcommand{\arraystretch}{0.92}
  \footnotesize
  \begin{tabular}{|l|c|c|c|c|}
    \hline
    $\sigma$ & $\hat{\sigma}^\star(w)[a(1)]$ & $\hat{\sigma}^\star(w)[a(4)]$ & $\hat{\sigma}^\star(w)[a(3)]$ & $\hat{\sigma}^\star(w)[a(2)]$ \\
    \hline
    $\hat{\sigma}^\star_{(1)}(w)$ & $1$ & $0$ & $0$ & $0$ \\
    \hline
    $\hat{\sigma}^\star_{(2)}(w)$ & $0$ & $1$ & $0$ & $0$ \\
    \hline
    $\hat{\sigma}^\star_{(3)}(w)$ &
      $\frac{1}{(1+\alpha)(1+\beta)}$ &
      $\frac{\alpha\beta}{(1+\alpha)(1+\beta)}$ &
      $\frac{\beta}{(1+\alpha)(1+\beta)}$ &
      $\frac{\alpha}{(1+\alpha)(1+\beta)}$ \\
    \hline
    $\hat{\sigma}^\star_{(4)}(w)$ &
      $\frac{1}{1+\beta+\alpha\beta}$ &
      $\frac{\alpha\beta}{1+\beta+\alpha\beta}$ &
      $\frac{\beta}{1+\beta+\alpha\beta}$ &
      $0$ \\
    \hline
    $\hat{\sigma}^\star_{(5)}(w)$ &
      $\frac{1}{1+\alpha+\alpha\beta}$ &
      $\frac{\alpha\beta}{1+\alpha+\alpha\beta}$ &
      $0$ &
      $\frac{\alpha}{1+\alpha+\alpha\beta}$ \\
    \hline
  \end{tabular}
  \endgroup
\end{table}

\begin{rem}[see~\cite{CalvoArmengol2003_CE2x2}]
For an \emph{anti}-coordination game with the reverse strict inequalities
$u^{w_1}_1(a(1)) < u^{w_1}_1(a(3))$, 
$u^{w_1}_1(a(4)) < u^{w_1}_1(a(2))$,
$u^{w_2}_2(a(1)) < u^{w_2}_2(a(2))$,
$u^{w_2}_2(a(4)) < u^{w_2}_2(a(3))$
the vertices in Table~\ref{tab:ce-vertices} are obtained by replacing \(\beta^{w_2}\) with \(1/\beta^{w_2}\) and swapping off-diagonal with diagonal probabilities.
In the dominance case, the sets of correlated and Nash equilibria coincide and collapse to a single point.
\end{rem}

In the coordination game $\Gamma^w
_2$, according to Theorem~\ref{th:1},
every element from the set of CE can be expressed by
\begin{align}\label{eq:CE_app1}
\hat{\sigma}^\star(w,y) =\sum_{v=1}^5 y_v\hat{\sigma}^\star_{(v)}(w),    
\end{align}
for some $y\in\Delta_5 =\{(y_1,\dots,y_5)\,:\, y_j\ge 0,\,\,  \sum_{j=1}^5 y_j = 1\}$.


Next, we assume the players in $\Gamma^w_2$ follow some correlated equilibrium and \(T\) independent observations \(\mathcal D = \{a^{(t)}=(a_1^{(t)},a_2^{(t)})\in A\}_{t=1}^T\) of the corresponding joint actions are provided.

Then, using the parametrized expression for the correlated equilibrium $\hat\sigma^\star(w,y) = (\hat\sigma^\star(w,y)[a(1)],\allowbreak
\hat\sigma^\star(w,y)[a(4)],\allowbreak
\hat\sigma^\star(w,y)[a(3)],\allowbreak
\hat\sigma^\star(w,y)[a(2)])$ as in~\eqref{eq:CE_app1}, we calculate the probability of the realized profile for each observation and cumulate them into the likelihood function:
$L_T(w,y)=\prod_{t=1}^{T}\hat\sigma^\star(w,y)[a^{(t)}]$.
In the last expression, $\hat\sigma^\star(w,y)[a^{(t)}]$ is equal to one of $\hat\sigma^\star(w,y)[a(l)]$, if the observation $a^{(t)}$ is the corresponding joint action $a(l)$.
The maximum-likelihood estimate solves the following problem: 
\begin{align}\label{eq:optCE}
    \min_{w,y}\{-\log(L_T(w,y)) =&\ -\sum_{t=1}^{T}\log\bigl(\hat\sigma^\star(w,y)[a^{(t)}]\bigr)\},\cr  &\mbox{s.t. }y\in\Delta_5. 
\end{align}
Thus, the objective of the optimization problem above is to find the best possible weights $y_v,\ v=1,\ldots,5$, and the utility parameters $w$ that jointly describe the empirical action frequencies in the dataset $\mathcal{D}$ through the equilibrium strategy $\hat{\sigma}^\star(w,y)$, i.e., to identify the parameter configuration that makes the observed actions most likely under the assumed game-theoretic behavioral model.
Hence, the part $w^*$ of the solution $(w^*, y^*)$ to~\eqref{eq:optCE} \textbf{solves the inverse game-theoretic learning} problem in $\Gamma^w_2$, given that the players choose a CE as a solution concept for their strategic interaction.

\subsection{Approach 2: Logit Best Response Learning (LBR-ML)}
 Correlated equilibrium-based formulations such as CE-ML presented in the previous subsection assume the observed behavior is influenced by a correlation device to generate the corresponding stable distribution. By contrast, the LBR-ML approach in this subsection models boundedly rational agents who iteratively adjust to one another over repeated interactions, formalized via the stationary state of the logit-best-response dynamics introduced in Section~\ref{sec:equilibrium-concepts}.

We consider again the $2\times 2$ game $\Gamma^w_2$ (see Table~\ref{tab:2x2-payoffs}) with $A_i=\{a_i^1,a_i^2\}$, $i=1,2$ and the joint action set $A=A_1\times A_2$ which, as before, is defined as follows: 
$A=\{a(1) = (a_1^1,a_2^1), a(2) = (a_1^1,a_2^2), a(3) = (a_1^2,a_2^1),a(4) = (a_1^2,a_2^2)\}$. 
Now we assume the players follow simultaneously the logit best response dynamics with finite positive $\lambda_1$ and $\lambda_2$ (see~\eqref{eq:lbr2}). Thus, the induced process is a four-state Markov chain with transition matrix
$P(\lambda,w)=(p_{kl}(\lambda, w))_{k,l=1}^4$.
According to Theorem~\ref{th:st}, for finite~$\lambda_i$, the dynamics converges to a unique stationary distribution  $\sigma^{\star}(\lambda,w)$ over $A$ satisfying~\eqref{eq:stationary}. This distribution is assumed to be the equilibrium solution concept in the LBR-ML approach. We proceed with estimation of $\sigma^{\star}(\lambda,w)$ in $\Gamma_2^w$ under consideration.

Let us define the one-step logit response probabilities using~\eqref{eq:lbr2} in $\Gamma_2^w$:
\[
\begin{aligned}
s_1(\lambda_1,w_1)=\sigma_1(\lambda_1,w_1)[a_1^1|{a_2^1}],\\
s_2(\lambda_1,w_1)=\sigma_1(\lambda_1,w_1)[a_1^1|{a_2^2}],\\
t_1(\lambda_2,w_2)=\sigma_2(\lambda_2,w_2)[a_2^1|{a_1^1}],\\
t_2(\lambda_2,w_2)=\sigma_2(\lambda_2,w_2)[a_2^1|{a_1^2}].
\end{aligned}
\]
Then, the transition matrix is defined as follows\footnote{We omit dependence in $s_1, s_2, t_1, t_2$ on $w$ and $\lambda$ for the sake of notation simplicity.}: 
\begin{align}\label{eq:Pm}
&P(\lambda,w)=(p_{kl}(\lambda, w))_{k,l=1}^4\cr
&\,=\resizebox{0.75\columnwidth}{!}{$\begin{bmatrix}
s_1 t_1 & s_1 (1-t_1) & (1-s_1) t_1 & (1-s_1)(1-t_1)\\[2pt]
s_2 t_1 & s_2 (1-t_1) & (1-s_2) t_1 & (1-s_2)(1-t_1)\\[2pt]
s_1 t_2 & s_1 (1-t_2) & (1-s_1) t_2 & (1-s_1)(1-t_2)\\[2pt]
s_2 t_2 & s_2 (1-t_2) & (1-s_2) t_2 & (1-s_2)(1-t_2)
\end{bmatrix}
$}.
\end{align}
Theorem~\ref{th:st}  implies existence and uniqueness of the solution to the system 
\[ \sigma(\lambda, w)\, P(\lambda,w) = \sigma(\lambda, w), \quad \mbox{s.t. }\, \sigma(\lambda, w)\boldsymbol{1} = 1.\]
In the $2\times 2$ game $\Gamma_2^w$ under consideration, the closed-form row-vector solution to the linear system above, i.e. the stationary distribution $\sigma^{\star}(\lambda, w)$, can be found as follows.
Due to the stationarity, we obtain the following affine system in respect to the marginal distributions of the players' actions:
\begin{align*}
\sigma_1^{\star}(\lambda, w)[a_1^1]=\sigma_2^{\star}(\lambda, w)[a_2^1]\,s_1+(1-\sigma_2^{\star}(\lambda, w)[a_2^1])s_2\cr
    \sigma_2^{\star}(\lambda, w)[a_2^1]=\sigma_1^{\star}(\lambda, w)[a_1^1]\,t_1+(1-\sigma_1^{\star}(\lambda, w)[a_1^1])t_2.
\end{align*}
By expressing $\sigma_1^{\star}(\lambda, w)[a_1^1]$ and $\sigma_1^{\star}(\lambda, w)[a_2^1]$ from the inequalities above, we obtain
$\sigma_1^{\star}(\lambda, w)[a_1^1]=\frac{s_2+(s_1-s_2)\,t_2}{1-(s_1-s_2)(t_1-t_2)}$ and
$\sigma_2^{\star}(\lambda, w)[a_2^1]=\frac{t_2+(t_1-t_2)\,s_2}{1-(s_1-s_2)(t_1-t_2)} $. We note that $1-(s_1-s_2)(t_1-t_2)\ne 0$ as we assume that the values $\lambda_1$ and $\lambda_2$ are finite (see again~\eqref{eq:lbr2}).  
Since we deal with the case of $2\times2$ normal-form game, the final stationary distribution can be formed from the marginal ones as follows:
\begin{align}\label{eq:final2}
\sigma^*(\lambda, w)[a(1)]&=xy, \cr {\sigma^*(\lambda, w)}[a(2)]&=x(1-y),\cr
{\sigma^*(\lambda, w)}[a(3)]&=(1-x)y, \cr {\sigma^*(\lambda, w)}[a(4)]&=(1-x)(1-y),
\end{align}
 where we use the notation $x = \sigma_1^{\star}(\lambda, w)[a_1^1]$, $y = \sigma_2^{\star}(\lambda, w)[a_2^1]$.

Now we assume the players' logit best response dynamics in $\Gamma^w_2$ is stabilized and \(T\) independent observations \(\mathcal D = \{a^{(t)}\in A \}_{t=1}^T\) of the corresponding joint actions are provided.
Then, analogously to the previous subsection, 
we can
calculate the probability of the realized profile for each observation based on the expressions in~\eqref{eq:final2}. The cumulative likelihood function is, thus, as follows: $
L_T(\lambda, w)=\prod_{t=1}^{T}\sigma^{\star}(\lambda,w)[a^{(t)}]$.
In the expression above, $\sigma^{\star}(\lambda, w)[a^{(t)}]$ is equal to  $\sigma^{\star}(\lambda, w)[a(l)]$, if the observation $a^{(t)}$ is the corresponding joint action $a(l)$. 
The maximum-likelihood estimate solves the following problem: 
\vspace{-0.1cm}
\begin{align}\label{eq:optLBR}
    \min_{\lambda,w}\{-L_T(\lambda, w) &= -\sum_{t=1}^{T}\log\bigl(\sigma^{\star}(\lambda,w)[a^{(t)}]\bigr)\},\cr 
    &\mbox{s.t. }\lambda_1,\lambda_2\ge 0. 
\end{align}
Thus, the objective of the optimization problem above is to find the best possible players' rationality parameters $\lambda=(\lambda_1,\lambda_2)$ and the utility weights $w$ that jointly reproduce the empirical joint-action frequencies observed in the dataset~$\mathcal{D}$ through the stationary equilibrium distribution $\sigma^{\star}(\lambda,w)$. The problem intends to identify the parameters that make the observed behavior most probable under the logit best-response dynamics.
In particular, the part $w^*$ of the solution $(\lambda^*, w^*)$ to~\eqref{eq:optLBR} \textbf{solves the inverse game-theoretic learning} problem in $\Gamma^w_2$, given that the players choose a stationary distribution of the LBR dynamics as a solution concept for their strategic interaction.

\section{Experimental Setup}

We assess our methods on a synthetic $2{\times}2$ game and a SUMO traffic scenario, moving from controlled to realistic conditions. In each case we specify a ground-truth game, generate data via a CE or simulation protocol, and fit ICE, CE-ML, and LBR-ML, reporting parameter recovery and distributional fit.


\subsection{Experimental Domains}

\subsubsection{Synthetic ``Chicken-Dare'' Game}
We adopt the canonical $2{\times}2$ Chicken-Dare interaction from \cite{bestick2013inverse}, where each player chooses \emph{Go} or \emph{Wait}. Utilities follow the linear specification in \eqref{eq:utility-linear} using the feature map and magnitudes from \cite{bestick2013inverse}; we omit the feature definitions here for brevity. For ground truth, we fix the players’ weights as defined for this experiment, compute the maximum-entropy correlated equilibrium (see~\cite{ME} for more details on this correlated equilibrium notion) in the resulting game, and draw i.i.d.\ joint actions from that CE. This isolates the inverse step by ensuring the data are CE-consistent for a known game.

\subsubsection{Traffic Interaction Game in SUMO}
\label{subsec:sumo}

To assess realism and applicability, we consider the same $2\times2$ interaction structure within a microscopic traffic simulator (SUMO \cite{SUMO2018}). This environment extends the abstract ``chicken-dare'' setup by introducing continuous kinematic dependencies and stochastic driver behavior, providing a realistic validation domain for the proposed inverse learning models.

\begin{figure}[ht]
\centering
\begin{tikzpicture}[scale=0.8, line cap=round, line join=round]

  \draw[line width=1.6pt] (-2.4,0.8) -- (-0.7,0.8) -- (-0.7,2.4);
  \draw[line width=1.6pt] ( 2.4,0.8) -- ( 0.7,0.8) -- ( 0.7,2.4);
  \draw[line width=1.6pt] (-2.4,-0.8) -- (-0.7,-0.8) -- (-0.7,-2.4);
  \draw[line width=1.6pt] ( 2.4,-0.8) -- ( 0.7,-0.8) -- ( 0.7,-2.4);

  \fill (0,0) circle (1.5pt);

  \def\tH{0.2}
  
  \def\sA{0.60} \def\yA{-1.55}
  \fill[blue!75!black] (0,\yA+\sA/2) -- (\sA/2,\yA-\tH+\sA/2) -- (\sA/2,\yA-\sA) -- (-\sA/2,\yA-\sA) -- (-\sA/2,\yA-\tH+\sA/2) -- cycle;
  \node[anchor=west] at (0.2,\yA-\sA/4) {$A$};

  \coordinate (Afront) at (0,\yA+\sA/2);

  \draw[-{Latex}, thick] (Afront) -- ++(0,0.75);
  \node[anchor=west] at (0,\yA+0.6) {$v_A$};

  \draw[line width=0.4pt] (-0.1,0) -- (-0.4,0);
  \draw[line width=0.4pt] (-0.4,\yA+\sA/2) -- (-0.1,\yA+\sA/2);
  \draw[dashed] (-0.35,0) -- (-0.35,\yA+\sA/2);
  \node[anchor=east] at (-0.35,-0.37) {$d_A$};

  \def\sB{0.60} \def\xB{1.55}
  \fill[orange!85!black] (\xB-\sB/2,0) -- (\xB+\tH-\sB/2,\sB/2) -- (\xB+\sB,\sB/2) -- (\xB+\sB,-\sB/2) -- (\xB+\tH-\sB/2,-\sB/2) -- cycle;
  \node[anchor=west] at (\xB+\sB,0) {$B$};

  \coordinate (Bfront) at (\xB-\sB/2,0);

  \draw[-{Latex}, thick] (Bfront) -- ++(-0.75,0);
  \node[anchor=north] at (\xB-0.6,-0.05) {$v_B$};

  \draw[line width=0.4pt] (0,0.1) -- (0,0.4);
  \draw[line width=0.4pt] (\xB-\sB/2,0.1) -- (\xB-\sB/2,0.4);
  \draw[dashed] (0,0.35) -- (\xB-\sB/2,0.35);
  \node[anchor=north] at (0.35,0.9) {$d_B$};

\end{tikzpicture}
\caption{Uncontrolled intersection used in SUMO. Distances $d_A,d_B$ and velocities $v_A,v_B$ define times $\tau_i=d_i/v_i$ and offset $\delta=\tau_2-\tau_1$.}
\label{fig:intersection}
\end{figure}
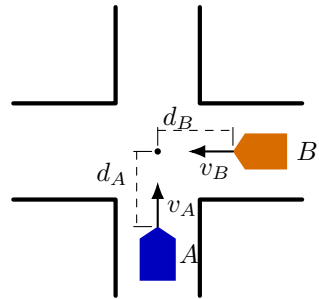

\paragraph{Scenario design}
Two agents (A and B) approach an uncontrolled intersection from orthogonal directions and choose $A_i=\{\text{wait},\text{go}\}$. Let $v_i$ be speed, $d_i$ distance to the conflict point, $\tau_i=d_i/v_i$ time to intersection, and $\delta=\tau_2-\tau_1$. Figure~\ref{fig:intersection} illustrates the geometry used to extract $(v_i,d_i)$ and derive $(\tau_i,\delta)$.

\paragraph{Kinematic-dependent utilities}

Utilities $u_i^{w_i}(a)$ follow the linear model in \eqref{eq:utility-linear}, where the feature vector $\phi_i(a;\tau_i,\tau_{-i},\delta)$ encodes interpretable \emph{kinematic} quantities that capture urgency ($\tau_i^{-1}$), time pressure ($\tau_i$), near-collision risk ($( |\delta|{+}\varepsilon )^{-1}$), and baseline preferences; the resulting payoff table is as follows:
\begin{equation}
\label{eq:traffic_payoff_matrix}
\renewcommand{\arraystretch}{1.2}
\resizebox{\linewidth}{!}{$
\begin{array}{@{}c|cc@{}}
 & a_2^1 & a_2^2 \\ \hline
a_1^1 &
\big(-w_{1,7}|\delta|{+}w_{1,8},\; -w_{2,7}|\delta|{+}w_{2,8}\big) &
\big(-w_{1,3}\tau_1{+}w_{1,4},\; \tfrac{w_{2,1}}{\tau_2}{+}w_{2,2}\big) \\
a_1^2 &
\big(\tfrac{w_{1,1}}{\tau_1}{+}w_{1,2},\; -w_{2,3}\tau_2{+}w_{2,4}\big) &
\big(-\tfrac{w_{1,5}}{|\delta|{+}\varepsilon}{+}w_{1,6},\;
 -\tfrac{w_{2,5}}{|\delta|{+}\varepsilon}{+}w_{2,6}\big)
\end{array}
$}
\end{equation}

Note that $\varepsilon = 10^{-6}$ is introduced to avoid potential division by zero. This specification follows risk/time formulations in traffic interaction models \cite{ZhangFricker2021ConflictRisksPMI} and alternative payoffs used for unsignalized crossings \cite{kalantari2023driver}; see also broader AV decision-making \cite{liu2022three}. 

\paragraph{Dataset extraction}
We sweep initial conditions in SUMO \cite{SUMO2018} and run each episode until the \emph{decision point} where agents choose their action $a_i\in \{\text{wait},\text{go}\}$. At that instant we log $(v_i,d_i)$, compute $(\tau_i,\delta)$, and extract the realized joint action $a\in A_1\times A_2$. Each datum is $\big((a_1,a_2),\phi_1(a;\tau_1,\tau_2,\delta),\phi_2(a;\tau_2,\tau_1,\delta)\big)$. 

\subsection{Compared Frameworks}
All experiments compare the following inverse game-theoretic estimators:
\begin{itemize}
    \item \textbf{ICE}~\cite{bestick2013inverse}: a convex program that enforces empirical correlated-equilibrium consistency.
    \item \textbf{CE-ML}: a maximum-likelihood estimator that fits a parametric correlated-equilibrium mixture 
    $\hat{\sigma}^\star(w,y)=\sum_{v=1}^{5}y_v\,\hat{\sigma}^\star_{(v)}(w)$, where $y\!\in\!\Delta_5$ are vertex weights (see optimization problem~\eqref{eq:optCE}).
    \item \textbf{LBR-ML}: a logit-best-response model estimating player utilities $w$ and rationality parameters 
    $(\lambda_1,\lambda_2)$ that define a stationary distribution of the induced Markov process (see optimization problem~\eqref{eq:optLBR}).
\end{itemize}





\subsection{Evaluation Metrics}
We report two complementary measures:

\begin{itemize}
  \item \textbf{Parameter recovery:} mean absolute error (MAE) and root-mean-square error (RMSE) between estimated and true utility weights $w_i^{\star}$.
  \item \textbf{Distributional fit:} empirical vs. predicted joint action probabilities in the corresponding equilibrium solution and decision accuracy (percentage of correctly predicted actions).
\end{itemize}

\subsection{Experiment Design}
\label{ssec:exp_design}

\textit{E1: Chicken via CE.}

\textbf{Purpose:} Benchmark each method on a fully known $2{\times}2$ game to verify parameter-recovery capability and distributional fit.\\
\textbf{Setup:} Use the synthetic ``Chicken--Dare'' game with ground-truth weights 
$w_1^\star = [\,0.3,\,0.7\,]^\top$ and $w_2^\star = [\,0.4,\,0.6\,]^\top$. 
We generate $T{\in}\{500,1000,2000\}$ samples from the maximum-entropy correlated equilibrium of this game.

\vspace{1mm}
\textit{E2: Traffic (MaxEnt-CE Ground Truth).}
\label{ssec:e2}

\textbf{Purpose:} Evaluate learning performance when traffic decisions follow an exact CE distribution.\\
\textbf{Setup:} Use the SUMO-based traffic game with ground-truth utility parameters $w^\star$ sampled from the maximum-entropy correlated equilibrium:
\begin{align}
\label{eq:e2_ground_truth}
w^{\star}_1 &= [\,0.05,\; 0.00,\; 0.02,\; 0.00,\; 0.90,\; 0.00,\; 0.03,\; 0.00\,], \nonumber\\
w^{\star}_2 &= [\,0.01,\; 0.00,\; 0.04,\; 0.00,\; 0.94,\; 0.00,\; 0.01,\; 0.00\,].
\end{align}

\textit{E3: Traffic with Signaled Correlation.}
\label{ssec:e3}

\textbf{Purpose:} Examine robustness when a correlating device generates partially coordinated behavior.\\
\textbf{Setup:} A signal selects joint actions $(a_1,a_2)\!\sim\!\sigma_{\mathrm{CE}}$ and privately sends recommendations to each driver, who follow their assigned actions. 

\vspace{1mm}
\textit{E4: Traffic without Coordination.}
\label{ssec:e4}

\textbf{Purpose:} Test the models on data not guaranteed to satisfy any equilibrium structure.\\
\textbf{Setup:} Simulate driver interactions in SUMO without any correlation mechanism; stochastic driver variability introduces behavioral noise. 

\section{Results}

The datasets provide a controlled benchmark to evaluate each framework’s ability to recover the ground-truth utility parameters $w_i^{\star}$ and reproduce observed joint-action distributions across four experiments (E1-E4). 
Fitted non-weight parameters $y$ and $\lambda$ from CE-ML and LBR-ML respectively are listed in Table \ref{tab:non_weight_parameters_exps} provided in Appendix.

\subsection*{E1. Chicken via CE}


The dataset $\mathcal{D}_\text{E1}=\{a^{(t)}\}^T_{t=1}$ provides an empirical distribution $\sigma_{\text{CE}}$ used by all methods, namely ICE, CE-ML, and LBR-ML, to estimate parameters ${w}_i^{\star}$ via maximum likelihood.

\begin{figure}[ht!]
    \centering
    \includegraphics[width=0.9\columnwidth]{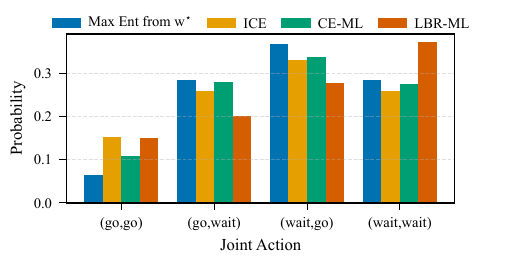}
    \caption{Probability distribution of equilibria per joint action, T = 2000.}
    \label{fig:exp1-dist}
\end{figure}

Table \ref{tab:exp1_mae_rmse_cells} and Figure \ref{fig:exp1-dist} summarize parameter recovery and predicted action probabilities.

While ICE and LBR-ML show limited improvement as the sample size $T$ increases, CE-ML shows clear improvement in accuracy and distributional fit, achieving the closest match to the ground-truth CE and outperforming the other methods.
For LBR-ML, this trend relates to the fitted rationality parameters $\lambda_i$ (see Table \ref{tab:non_weight_parameters_exps} in Appendix), which spike at $T{=}1000$. The higher $\lambda$-values imply more deterministic play, reducing stochastic smoothing and increasing sensitivity to sampling noise, consistent with the temporary rise in error at that horizon. As $T$ grows further, the $\lambda$-values fall again, yielding smoother behavior and improved fit.

\input{results/exp1/summary_wide}

\subsection*{E2: Traffic (MaxEnt-CE Ground Truth)}
This experiment evaluates the inverse models on the SUMO-based traffic game, using the data set 
$\mathcal{D}_{\mathrm{E2}}=\{a^{(t)}\}_{t=1}^{T}$ 
sampled from the maximum-entropy correlated equilibrium of the ground-truth parameters.



Parameter-recovery errors are shown in Table~\ref{tab:exp2_mae_rmse_cells}. CE-ML achieves the best overall performance, followed by LBR-ML and then ICE, an interesting result given that ICE assumes CE-consistent data, whereas LBR-ML relies on a different equilibrium notion. For LBR-ML, the fitted rationality parameters $\lambda_i$ remain consistently high across sample sizes (see Table \ref{tab:non_weight_parameters_exps} in Appendix), indicating that both agents behave in a near-deterministic manner. This stability suggests that the interaction is already close to a coordinated equilibrium structure, where increasing $T$ adds little new stochastic information for the model to learn.

\input{results/exp2/summary_wide}

\begin{figure}[ht!]
    \centering
    \includegraphics[width=0.9\columnwidth]{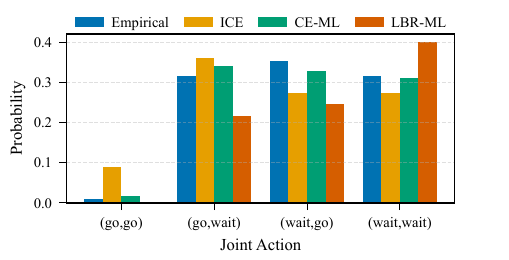}
    \caption{Probability distribution of equilibria per joint action, T = 2000.}
    \label{fig:exp2-dist}
\end{figure}

The corresponding joint action distributions in equilibria under consideration are shown in Fig.~\ref{fig:exp2-dist}.
Among the three frameworks, CE-ML yields the distribution most consistent with the ground truth.

The results from $\mathcal{D}_\text{E2}$ confirm that all inverse formulations can recover consistent utilities when the observed data follow a true correlated equilibrium. To further assess their robustness under coordination, the next experiment introduces a signaling mechanism that mediates driver behavior.

\subsection*{E3: Traffic with Signaled Correlation}

A correlating device generates joint actions $(a_1,a_2)\!\sim\!\sigma_{\mathrm{CE}}$ and privately recommends actions to each player, who follow their signals. 
The resulting dataset $\mathcal{D}_\text{E3}=\{a^{(t)}\}_{t=1}^{500}$ captures fully CE-consistent, signal-mediated interactions. 
This experiment compares ICE and CE-ML in their ability to recover the underlying utilities and reproduce the induced correlated behavior.

\begin{figure}[ht!]
    \centering
    \includegraphics[width=0.9\columnwidth]{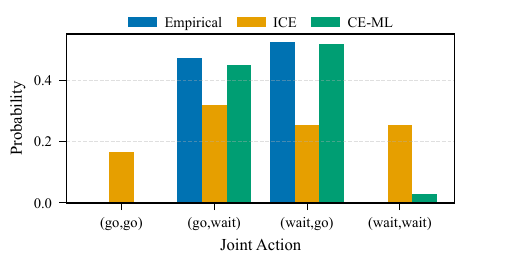}
    \caption{Probability distribution of equilibria per joint action, T=500}
    \label{fig:exp3-dist}
\end{figure}

Accuracy results are summarized in Table~\ref{tab:exp3_accuracy_summary}, and the corresponding joint-action distributions are shown in Fig.~\ref{fig:exp3-dist}.
CE-ML achieves the best alignment with the observed data and lowest parameter error, as expected under CE-consistent observations, while ICE display comparatively weaker fit.

\input{results/exp3/table_accuracy_summary}

The estimated CE-ML mixture weights $y_i$ remain stable across sample sizes and concentrate on the same equilibrium components that generate the data, indicating that CE-ML recovers the structural correlation pattern induced by the signaling device (see Table~\ref{tab:non_weight_parameters_exps} in Appendix).

\subsection*{E4: Traffic without Coordination}

This final experiment examines a setting where no explicit coordination or signaling mechanism is present.
Each driver acts independently in the SUMO simulation, producing a dataset $\mathcal{D}_\text{E4}=\{a^{(t)}\}^{500}_{t=1}$ that reflects naturally emerging, potentially non-equilibrium behavior.
The absence of a correlation device makes this scenario particularly relevant for evaluating model robustness under real-world stochasticity and suboptimal decision patterns.

\begin{figure}[ht!]
    \centering
    \includegraphics[width=0.9\columnwidth]{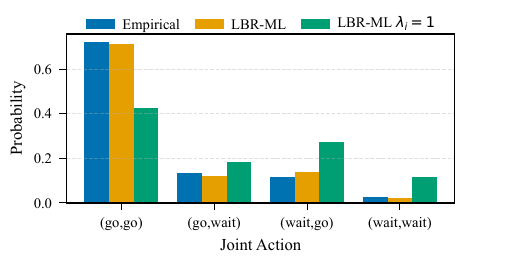}
    \caption{Probability distribution of equilibria per joint action, T=500}
    \label{fig:exp4-dist}
\end{figure}

In this  setting, CE-ML and ICE no longer align well with the empirical behavior, as their equilibrium assumptions are violated.
Therefore, we focus on LBR-ML, comparing the version with fixed $\lambda_i=1$ to the variant where both $\lambda_i$ are estimated from data (see Table \ref{tab:non_weight_parameters_exps} in Appendix).
Both configurations achieve the same \textbf{accuracy of 72.6$\%$}, yet the estimated version, with $\lambda_1{=}1.0$ and $\lambda_2{=}3.0$, produces a joint-action distribution that more closely matches the empirical frequencies (Fig.~\ref{fig:exp4-dist}).
This result indicates that allowing $\lambda_i$ to adapt captures agents’ heterogeneous bounded rationality, yielding a more flexible and behaviorally consistent representation of non-equilibrium interactions, even in the absence of coordination or signaling.

Across the four experiments, the proposed inverse game-theoretic frameworks exhibit consistent trends. 
CE-ML performs best when the observed behavior aligns with correlated equilibrium assumptions (E1--E3), while LBR-ML turns to be more robust under uncoordinated or noisy conditions (E4). 
ICE offers a general baseline but shows limited improvement with data scale or deviation from CE-consistent settings. 
The evolution of the fitted rationality parameters $(\lambda_1,\lambda_2)$ across experiments further supports these findings: higher $\lambda$ values emerge in structured, coordinated settings, reflecting near-deterministic play under equilibrium-like behavior, whereas lower or heterogeneous $\lambda$ values appear in adaptive, non-equilibrium scenarios such as E4. 
These patterns highlight the complementary strengths of the proposed methods: CE-ML for structured coordination, and LBR-ML for capturing varying levels of bounded rationality and adaptive decision-making in environments without any correlation device.

\section{Conclusion}
This work solves distinct inverse game-theoretic formulations to capture different aspects of interactive behavior in the case of $2\times 2$ normal-form games. The proposed CE-ML approach provides accurate and computationally efficient parameter recovery when the data are consistent with a correlated-equilibrium structure, outperforming the existing ICE method from~\cite{bestick2013inverse} in CE-generated or signal-mediated scenarios.
The second approach, LBR-ML, by contrast, better accounts for adaptive interactions under bounded rationality maintaining predictive power even when coordination assumptions break down. Future work will extend these frameworks to multi-vehicle and heterogeneous-agent settings, and explore online estimation of utility parameters from real or partially observed traffic interactions.


\section*{Appendix}

\input{results/non_weight_parameters_all_exp}

\bibliographystyle{IEEEtran}
\bibliography{references}

\end{document}

%% file: results/exp1/summary_wide.tex
\begin{table}[ht!]
\centering
\caption{Aggregate MAE/RMSE over all weights versus sample size $T$. Lower is better; bold marks the best MAE at each $T$.}
\begin{tabular}{lccc}
\toprule
Model & $T\,=\,500$ & $T\,=\,1000$ & $T\,=\,2000$ \\
\midrule
ICE & 0.133 / 0.139 & 0.133 / 0.137 & 0.126 / 0.131 \\
CE-ML & \textbf{0.133 / 0.138} & \textbf{0.114 / 0.114} & \textbf{0.084 / 0.097} \\
LBR-ML & 0.155 / 0.194 & 0.433 / 0.442 & 0.131 / 0.176 \\
\bottomrule
\end{tabular}
\label{tab:exp1_mae_rmse_cells}
\end{table}

%% file: results/exp2/summary_wide.tex
\begin{table}[ht!]
\centering
\caption{Aggregate MAE/RMSE over all weights versus sample size $T$. Lower is better; bold marks the best MAE at each $T$.}
\begin{tabular}{lccc}
\toprule
Model & $T\,=\,500$ & $T\,=\,1000$ & $T\,=\,2000$ \\
\midrule
ICE & 0.111 / 0.143 & 0.111 / 0.144 & 0.111 / 0.144 \\
CE-ML & \textbf{0.019 / 0.023} & \textbf{0.017 / 0.022} & \textbf{0.017 / 0.022} \\
LBR-ML & 0.041 / 0.050 & 0.040 / 0.049 & 0.043 / 0.051 \\
\bottomrule
\end{tabular}
\label{tab:exp2_mae_rmse_cells}
\end{table}

%% file: results/exp3/table_accuracy_summary.tex
\begin{table}[ht!]
\centering
\caption{Accuracy across SUMO configurations. Bold indicates the best-performing model.}
\begin{tabular}{lc}
\toprule
Model & Accuracy (\%) \\
\midrule
ICE & 41.80\% \\
CE-ML & \textbf{86.40}\% \\
\bottomrule
\end{tabular}
\label{tab:exp3_accuracy_summary}
\end{table}

%% file: results/non_weight_parameters_all_exp.tex
\begin{table}[ht!]
\centering
\caption{Non-weight parameters by experiment and sample $T$. CE-ML mixture weights $y_i$ to CE vertices $v_i$.}
\setlength{\tabcolsep}{4pt}
\begin{tabular}{l c ccccc cc}
\toprule
 & & \multicolumn{5}{c}{CE-ML ($y_i$)} & \multicolumn{2}{c}{LBR-ML ($\lambda_i$)} \\
\cmidrule(lr){3-7} \cmidrule(l){8-9}
Exp. & $T$ & $y_1$ & $y_2$ & $y_3$ & $y_4$ & $y_5$ & $\lambda_1$ & $\lambda_2$ \\
\midrule
E1 & 500 & 0.09 & 0.10 & 0.63 & 0.04 & 0.14 & 0.50 & 0.50 \\
E1 & 1000 & 0.14 & 0.01 & 0.66 & 0.02 & 0.17 & 2.00 & 2.00 \\
E1 & 2000 & 0.00 & 0.05 & 0.52 & 0.00 & 0.43 & 0.50 & 1.00 \\
\midrule
E2 & 500 & 0.02 & 0.10 & 0.00 & 0.36 & 0.52 & 3.00 & 3.00 \\
E2 & 1000 & 0.04 & 0.10 & 0.00 & 0.36 & 0.51 & 3.00 & 3.00 \\
E2 & 2000 & 0.00 & 0.07 & 0.00 & 0.36 & 0.57 & 3.00 & 3.00 \\
\midrule
E3 & 500 & 0.00 & 0.00 & 0.00 & 1.00 & 0.00 & -- & -- \\
\midrule
E4 & 500 & 0.00 & 0.00 & 1.00 & 0.00 & 0.00 & 1.00 & 3.00 \\
\bottomrule
\label{tab:non_weight_parameters_exps}
\end{tabular}
\end{table}

%% file: references.bib
@inproceedings{SUMO2018,
          title = {Microscopic Traffic Simulation using SUMO},
         author = {Pablo Alvarez Lopez and Michael Behrisch and Laura Bieker-Walz and Jakob Erdmann and Yun-Pang Fl{\"o}tter{\"o}d and Robert Hilbrich and Leonhard L{\"u}cken and Johannes Rummel and Peter Wagner and Evamarie Wie{\ss}ner},
      publisher = {IEEE},
      booktitle = {The 21st IEEE International Conference on Intelligent Transportation Systems},
           year = {2018},
        journal = {IEEE Intelligent Transportation Systems Conference (ITSC)},
            url = {https://elib.dlr.de/124092/}
 }

@article{CostaGomesCrawfordBroseta2001,
  author    = {Costa-Gomes, Miguel A. and Crawford, Vincent P. and Broseta, Bruno},
  title     = {Cognition and Behavior in Normal-Form Games: An Experimental Study},
  journal   = {Econometrica},
  volume    = {69},
  number    = {5},
  pages     = {1193--1235},
  year      = {2001},
  doi       = {10.1111/1468-0262.00247}
}

@article{RazinFeigh+2021+481+502,
url = {https://doi.org/10.1515/pjbr-2021-0031},
title = {Committing to interdependence: Implications from game theory for human–robot trust},
author = {Yosef S. Razin and Karen M. Feigh},
pages = {481--502},
volume = {12},
number = {1},
journal = {Paladyn, Journal of Behavioral Robotics},
doi = {doi:10.1515/pjbr-2021-0031},
year = {2021},
lastchecked = {2025-10-28}
}

@article{PapadimitriouRoughgarden2008,
  author    = {Papadimitriou, Christos H. and Roughgarden, Tim},
  title     = {Computing Correlated Equilibria in Multi-Player Games},
  journal   = {Journal of the ACM},
  volume    = {55},
  number    = {3},
  pages     = {14:1--14:29},
  year      = {2008},
  doi       = {10.1145/1379759.1379762}
}

@book{Tatarenko2017GameTheoreticLearning,
  author       = {Tatarenko, Tatiana},
  title        = {Game-Theoretic Learning and Distributed Optimization in Memoryless Multi-Agent Systems},
  publisher    = {Springer International Publishing},
  address      = {Cham, Switzerland},
  year         = {2017},
  isbn         = {978-3-319-65478-2},
  doi          = {10.1007/978-3-319-65479-9}
}

@book{Seneta2006,
  author    = {Seneta, Eugene},
  title     = {Non-negative Matrices and Markov Chains},
  publisher = {Springer},
  address   = {New York},
  edition   = {revised},
  year      = {2006},
  doi       = {10.1007/0-387-29765-7}
}

@article{Blume1993,
  author    = {Blume, L.\,E.},
  title     = {The statistical mechanics of strategic interaction},
  journal   = {Games and Economic Behavior},
  volume    = {5},
  number    = {3},
  pages     = {387--424},
  year      = {1993},
  doi       = {10.1006/game.1993.1020}
}

@article{ZhangFricker2021ConflictRisksPMI,
  author  = {Zhang, Yunchang and Fricker, Jon D.},
  title   = {Incorporating Conflict Risks in Pedestrian–Motorist Interactions: A Game Theoretical Approach},
  journal = {Accident Analysis \& Prevention},
  year    = {2021},
  volume  = {159},
  pages   = {106254},
  doi     = {10.1016/j.aap.2021.106254}
}

@article{ArbisDixit2019LaneChangingConflicts,
  author  = {Arbis, D. and Dixit, V. V.},
  title   = {Game theoretic model for lane changing: Incorporating conflict risks},
  journal = {Accident Analysis \& Prevention},
  year    = {2019},
  volume  = {125},
  pages   = {158--164},
  doi     = {10.1016/j.aap.2019.02.007}
}

@article{DuffyLaiLim2017CoordinationViaCorrelation,
  author  = {Duffy, John and Lai, Ernest K. and Lim, Wooyoung},
  title   = {Coordination via Correlation: An Experimental Study},
  journal = {Economic Theory},
  year    = {2017},
  volume  = {64},
  number  = {2},
  pages   = {265--304},
  doi     = {10.1007/s00199-016-0970-y}
}

@InProceedings{pmlr-v97-yu19e,
  title     = {Multi-Agent Adversarial Inverse Reinforcement Learning},
  author    = {Yu, Lantao and Song, Jiaming and Ermon, Stefano},
  booktitle = {Proceedings of the 36th International Conference on Machine Learning},
  pages     = {7194--7201},
  year      = {2019},
  editor    = {Chaudhuri, Kamalika and Salakhutdinov, Ruslan},
  volume    = {97},
  series    = {Proceedings of Machine Learning Research},
  month     = {09--15 Jun},
  publisher = {PMLR},
  url       = {https://proceedings.mlr.press/v97/yu19e.html}
}

@inproceedings{KuleshovSchrijvers2015InverseGameTheory,
  author    = {Volodymyr Kuleshov and Okke Schrijvers},
  title     = {Inverse Game Theory: Learning Utilities in Succinct Games},
  booktitle = {Web and Internet Economics: 11th International Conference, WINE 2015, Amsterdam, The Netherlands, December 9--12, 2015, Proceedings},
  editor    = {Evangelos Markakis and Guido Sch{\"a}fer},
  series    = {Lecture Notes in Computer Science},
  volume    = {9470},
  pages     = {413--427},
  publisher = {Springer},
  address   = {Cham},
  year      = {2015},
  doi       = {10.1007/978-3-662-48995-6_30},
  url       = {https://link.springer.com/chapter/10.1007/978-3-662-48995-6_30}
}

@inproceedings{Chandra2025MAIRLcrowdsIROS,
  title        = {Multi-Agent Inverse Reinforcement Learning in Real World Unstructured Pedestrian Crowds},
  author       = {Chandra, Rohan and Karnan, Haresh and Mehr, Negar and Stone, Peter and Biswas, Joydeep},
  booktitle    = {Proceedings of the IEEE/RSJ International Conference on Intelligent Robots and Systems (IROS)},
  year         = {2025},
  note         = {to appear}
}

@inproceedings{bestick2013inverse,
  title={An inverse correlated equilibrium framework for utility learning in multiplayer, noncooperative settings},
  author={Bestick, Aaron and Ratliff, Lillian J and Yan, Posu and Bajcsy, Ruzena and Sastry, S Shankar},
  booktitle={Proceedings of the 2nd ACM international conference on High confidence networked systems},
  pages={9--16},
  year={2013}
}

@article{MCKELVEY19956,
title = {Quantal Response Equilibria for Normal Form Games},
journal = {Games and Economic Behavior},
volume = {10},
number = {1},
pages = {6-38},
year = {1995},
issn = {0899-8256},
doi = {https://doi.org/10.1006/game.1995.1023},
url = {https://www.sciencedirect.com/science/article/pii/S0899825685710238},
author = {Richard D. McKelvey and Thomas R. Palfrey}
}

@techreport{CalvoArmengol2003_CE2x2,
  author      = {Antoni Calv{\'o}-Armengol},
  title       = {The Set of Correlated Equilibria of 2 × 2 Games},
  institution = {Barcelona Economics Working Paper Series},
  number      = {79},
  year        = {2003},
  month       = {October},
  address     = {Barcelona},
  url         = {https://bse.eu/sites/default/files/working_paper_pdfs/79.pdf}
}

@inproceedings{ME,
  title={Maximum entropy correlated equilibria},
  author={Ortiz, Luis E and Schapire, Robert E and Kakade, Sham M},
  booktitle={Artificial Intelligence and Statistics},
  pages={347--354},
  year={2007},
  organization={PMLR}
}

@article{hart2000simple,
  title={A simple adaptive procedure leading to correlated equilibrium},
  author={Hart, Sergiu and Mas-Colell, Andreu},
  journal={Econometrica},
  volume={68},
  number={5},
  pages={1127--1150},
  year={2000},
  publisher={Wiley Online Library}
}

@article{daskalakis2009complexity,
  title={The complexity of computing a Nash equilibrium},
  author={Daskalakis, Constantinos and Goldberg, Paul W and Papadimitriou, Christos H},
  journal={Communications of the ACM},
  volume={52},
  number={2},
  pages={89--97},
  year={2009},
  publisher={ACM New York, NY, USA}
}

@article{aumann1974subjectivity,
  title={Subjectivity and correlation in randomized strategies},
  author={Aumann, Robert J},
  journal={Journal of mathematical Economics},
  volume={1},
  number={1},
  pages={67--96},
  year={1974},
  publisher={Elsevier}
}

@book{camerer2003behavioral,
  title={Behavioral Game Theory: Experiments in Strategic Interaction},
  author={Camerer, Colin F.},
  year={2003},
  publisher={Princeton University Press}
}

@article{nyarko2002belief,
  title={An experimental study of belief learning using elicited beliefs},
  author={Nyarko, Yaw and Schotter, Andrew},
  journal={Econometrica},
  volume={70},
  number={3},
  pages={971--1005},
  year={2002},
  publisher={Wiley Online Library}
}

@article{liu2022three,
  title={A three-level game-theoretic decision-making framework for autonomous vehicles},
  author={Liu, Mushuang and Wan, Yan and Lewis, Frank L and Nageshrao, Subramanya and Filev, Dimitar},
  journal={IEEE Transactions on Intelligent Transportation Systems},
  volume={23},
  number={11},
  pages={20298--20308},
  year={2022},
  publisher={IEEE}
}

@article{kalantari2023driver,
  title={Driver-pedestrian interactions at unsignalized crossings are not in line with the Nash equilibrium},
  author={Kalantari, Amir Hossein and Yang, Yue and Lee, Yee Mun and Merat, Natasha and Markkula, Gustav},
  journal={IEEE Access},
  volume={11},
  pages={110707--110723},
  year={2023},
  publisher={IEEE}
}
